\documentclass[preprint,superscriptaddress,preprintnumbers,amsmath,amssymb,prd]{revtex4}
\usepackage{graphicx}

\begin{document}

\thispagestyle{empty}

\title{The Nernst heat theorem for an atom interacting with graphene: Dirac model
with nonzero energy gap and chemical potential
}

\author{
G.~L.~Klimchitskaya}
\affiliation{Central Astronomical Observatory at Pulkovo of the
Russian Academy of Sciences, Saint Petersburg,
196140, Russia}
\affiliation{Institute of Physics, Nanotechnology and
Telecommunications, Peter the Great Saint Petersburg
Polytechnic University, Saint Petersburg, 195251, Russia}

\author{
V.~M.~Mostepanenko}
\affiliation{Central Astronomical Observatory at Pulkovo of the
Russian Academy of Sciences, Saint Petersburg,
196140, Russia}
\affiliation{Institute of Physics, Nanotechnology and
Telecommunications, Peter the Great Saint Petersburg
Polytechnic University, Saint Petersburg, 195251, Russia}
\affiliation{Kazan Federal University, Kazan, 420008, Russia}

\begin{abstract}
We derive the low-temperature behavior of the Casimir-Polder free energy
for a polarizable atom interacting with graphene sheet which possesses the
nonzero energy gap $\Delta$ and chemical potential $\mu$. The response of
graphene to the electromagnetic field is described by means of the
polarization tensor in the framework of Dirac model on the basis of first
principles of thermal quantum field theory in the Matsubara formulation.
It is shown that the thermal correction to the Casimir-Polder energy
consists of three contributions. The first of them is determined by the
Matsubara summation using the polarization tensor defined at zero
temperature, whereas the second and third contributions are caused by an
explicit temperature dependence of the polarization tensor and originate
from the zero-frequency Matsubara term and the sum of all Matsubara terms
with nonzero frequencies, respectively. The asymptotic behavior for each
of the three contributions at low temperature is found analytically for
any value of the energy gap and chemical potential. According to our
results, the Nernst heat theorem for the Casimir-Polder free energy and
entropy is satisfied for both $\Delta > 2\mu$ and $\Delta < 2\mu$. We
also reveal an entropic anomaly arising in the case $\Delta = 2\mu$. The
obtained results are discussed in connection with the long-standing
fundamental problem in Casimir physics regarding the proper
description of the dielectric response of matter to the electromagnetic
field.
\end{abstract}

\maketitle\newcommand{\tp}{{\tilde{\Pi}}}
\newcommand{\rM}{{r_{\rm TM}}}
\newcommand{\rE}{{r_{\rm TE}}}
\newcommand{\orM}{{r_{\rm TM}^{(0)}}}
\newcommand{\orE}{{r_{\rm TE}^{(0)}}}
\newcommand{\zy}{{({\rm i}\zeta_l,y,T)}}
\newcommand{\ozy}{{({\rm i}\zeta_l,y,0)}}
\newcommand{\oyt}{{(0,y,T)}}
\newcommand{\ri}{{{\rm i}}}
\newcommand{\cF}{{\cal{F}}(a,T)}
\newcommand{\ocF}{{\cal{F}}}
\newcommand{\vF}{{\tilde{v}_F}}
\newcommand{\ho}{{\hbar\omega_c}}
\newcommand{\dT}{{\delta_T}}
\newcommand{\daT}{{\delta_T^{\rm impl}}}
\newcommand{\dbT}{{\delta_{T\!,\,l=0}^{\rm expl}\,}}
\newcommand{\dcT}{{\delta_{T\!,\,l\geqslant 1}^{\rm expl}\,}}

\newcommand{\yt}{{(y,T,\Delta,\mu)}}
\newcommand{\yo}{{(y,0,\Delta,\mu)}}
\newcommand{\eEt}{e^{\frac{t\Delta-2\mu}{2k_BT}}}
\newcommand{\deE}{e^{-\frac{\Delta-2\mu}{2k_BT}}}
\newcommand{\meE}{e^{-\frac{2\mu-\Delta}{2k_BT}}}

\section{Introduction}
At the moment there is a strong interest to theoretical and experimental
investigations of the Casimir \cite{1} and Casimir-Polder \cite{2} forces
which act between electrically neutral bodies spaced at short separations
one from the other. These forces are of entirely quantum nature and are
caused by the zero-point and thermal fluctuations of the electromagnetic
field. At separations exceeding several nanometers the Casimir force,
which acts between two macroscopic bodies, and the Casimir-Polder force,
acting between an atom and a material surface, are of relativistic
character by depending on both the Planck constant $\hbar$ and the speed
of light $c$. In fact these forces present the relativistic generalization
of the familiar van der Waals forces \cite{3}, but take on greater
significance due to multidisciplinary applications not only in atomic
physics \cite{4,5,6,7,8,9,10,11,12} and condensed matter physics
\cite{13,14,15}, but also in quantum field theory \cite{15a,16,16a},
gravitation and cosmology \cite{17,18,19,20}, and for constraining
predictions of high energy physics, supersymmetry and supergravity
\cite{21,22,25,26,27,28}.

Theoretical description of the Casimir and Casimir-Polder forces
between real material bodies is based
on the semiclassical Lifshitz theory \cite{15b,16b}, which treats the
electromagnetic field in the framework of quantum field theory, but uses
the classical description of matter by means of some phenomenological
response functions (in certain limits one can derive these forces
without involving the Lifshitz theory, see, e.g., Refs.~\cite{1,2,30aa}).
In the framework of this theory, the Casimir-Polder
atom-plate interaction is expressed via the frequency dependent atomic
polarizability and the dielectric permittivity of a plate material.
Although the Lifshitz theory was successfully used over a period of
several decades, modern precise experiments performed during the
last few years revealed serious contradictions between experiment and
theory. Specifically, for two metallic bodies the theoretical predictions
obtained with taken into account relaxation properties of free
(conduction) electrons were found to be in a
not so far reconcilable contradiction
with the measurement data
(see Refs.~\cite{17b,19b,21b,22b,23b,25b,26b,27b,28b}
and reviews in Refs.~\cite{29,30,31}). The
contradiction arises if the available optical data of a metal are
extrapolated down to zero frequency by the well tested Drude model taking
the proper account of the relaxation properties of free electrons and
dies away if the lossless plasma model is used which should be applicable
only at high frequencies.

By an intriguing coincidence, the Casimir entropy, calculated using the
Lifshitz theory combined with the Drude model, does not vanish with
vanishing temperature for metals with perfect crystal lattices and
depends on the volume and other parameters of a system
\cite{34,35,35a,36,37}.
Thus, the Nernst heat theorem, which demands that for a
physical system in thermal equilibrium the entropy at zero temperature
must either vanish or be equal to the universal constant independent
on the system parameters \cite{37a,37b}, is violated in this case. In
doing so, the Nernst heat theorem is satisfied if the plasma model is
used \cite{34,35,35a,36,37} which is consistent with measurements
of the Casimir force, but is in conflict with all our knowledge about
the electric phenomena occurring at low frequencies. It was noticed also
\cite{38,39,40} that the Casimir entropy jumps to zero at a very low
temperature starting from the negative value if the Drude model is used
for metals with an imperfect crystal lattice containing some fraction
of impurities. This observation, however, does not help to bring the
Drude-based theory in agreement with the measurement results for the Casimir
force.

Somewhat similar situation was discovered for the Casimir force between
two dielectric bodies and for the Casimir-Polder force between a
polarizable atom in close proximity to a dielectric plate. It was found
that theoretical predictions of the Lifshitz theory obtained with
taken into account conductivity at a constant current (dc conductivity)
of a dielectric material are in contradictions with the measurement data
of Casimir experiments \cite{42,43,44,45}. To bring the theoretical
predictions in agreement with the measurement data, one needs to omit in
computations the really observable dc conductivity of a material
\cite{11,42,43,44,45}. It seems meaningful that the calculated values
of both the Casimir and Casimir-Polder entropies at zero temperature were
found to violate the Nernst heat theorem if the dc conductivity of a
dielectric body is included in calculations and in agreement with this
theorem otherwise \cite{47,48,49,50,51}. Thus, the theoretical approach
consistent with the results of Casimir experiments, in spite of its
inconsistency with clearly established facts in other fields of physics,
was again found in accordance with the requirements of thermodynamics.
The above contradictions have often been called in the literature
the Casimir puzzle and the Casimir conundrum
(see, e.g., Refs.~\cite{51a,52,52a,52b,53})  which still remain unresolved.

{}From the above reasoning it may be suggested that the Nernst heat theorem
plays an important role as a test for different approaches to a description
of the dielectric response of matter. The weak point of existing
approaches is the use of phenomenological local dielectric permittivities
given by the Drude and plasma models. It is the matter of fact that real
dielectrics and metals are too complicated systems, so that their response
to the electromagnetic field cannot be found exactly on the basis of first
principles of thermal quantum field theory. In this regard, much attention
is currently attracted to graphene which is a 2D-sheet of carbon atoms
packed in a hexagonal lattice. The remarkable feature of graphene is that
at energies below 1--2~eV it is described by the Dirac model where the
speed of light is replaced with the Fermi velocity $v_F \approx c/300$
\cite{54,55,56}. This opens opportunities for a full description of the
nonlocal dielectric properties of graphene in the framework of thermal
quantum field theory in the Matsubara formulation. It should be noted
also that the Casimir-Polder interaction of different atoms with graphene
and graphene-coated substrates attracts much recent attention
\cite{57,58,59,60,61,62,63,64,65,66}. This raises a question on whether or not the
Casimir and Casimir-Polder entropy in graphene systems is consistent with
the Nernst heat theorem.

This question can be investigated by describing the dielectric response
of graphene in terms of its polarization tensor. The exact expressions for
the polarization tensor of graphene with a nonzero energy gap $\Delta$ at
zero temperature have been found in Ref.~\cite{67}. In Ref.~\cite{68} they
were generalized for the case of nonzero temperature, but only at the
pure imaginary Matsubara frequencies. In Ref. \cite{69} another
representation for the polarization tensor of graphene was obtained valid
over the entire plane of complex frequencies. In Ref.~\cite{70} it was
generalized to the case of nonzero chemical potential $\mu$. A validity
of the Kramers-Kronig relations for the obtained dielectric response has
been demonstrated in Ref.~\cite{71}. Thus, it was proven that the
dielectric response of graphene satisfies the causality condition. Using
the results of Ref.~\cite{69}, it was shown that the Casimir entropy of
two parallel sheets of pristine graphene, possessing the zero energy gap
and chemical potential, as well as the Casimir-Polder entropy for an atom
interacting with a pristine graphene sheet, satisfy the Nernst heat theorem
\cite{72,72a}. The low-temperature expansion of the Casimir-Polder free
energy for an atom interacting with real graphene sheet possessing any
values of $\Delta$ and $\mu$ was considered in Ref.~\cite{73}, and several
main terms under different relationships between $\Delta$ and $\mu$ have
been found. Some of them, however, turned out to be in disagreement with
the results of Ref.~\cite{74} obtained only in the special case
$\Delta >2\mu$. Thus, the issue on a validity of the Nernst heat theorem for an
atom interacting with real graphene sheet remained open.

In this paper, we investigate the analytic behavior of the Casimir-Polder
free energy and entropy at low temperature for an atom interacting with
real graphene sheet for any relationships between the energy gap $\Delta$
and chemical potential $\mu$ basing on first principles of thermal
quantum field theory in the Matsubara formulation. For this purpose,
the thermal correction to the Casimir-Polder energy is presented as
a sum of three contributions. The first of them is obtained using the
polarization tensor of graphene at zero temperature. In this case the
temperature dependence arises only due to a summation over the Matsubara
frequencies. The second and third contributions originate from an
explicit dependence of the polarization tensor on temperature as a
parameter in the Matsubara term with zero frequency and in the sum of
terms with all nonzero Matsubara frequencies, respectively. It is shown
that for $\Delta > 2\mu$ the Casimir-Polder free energy at sufficiently
low temperature behaves as $\sim(k_{B}T)^5$ where $k_{B}$ is the Boltzmann
constant whereas for $\Delta < 2\mu$ as $\sim(k_{B}T)^2$. These behaviors
are determined by the first contribution to the thermal correction. The
conclusion is made that for $\Delta > 2\mu$ and $\Delta < 2\mu$ the
Casimir-Polder free energy and entropy for an atom interacting with
graphene sheet are in agreement with the Nernst heat theorem. The main
terms of the second and third contributions in the thermal correction
to the Casimir-Polder energy are also found. According to the obtained
results, for $\Delta = 2\mu$ the Casimir-Polder free energy at low
temperature is of the order of $k_{B}T$ and is determined by the third
contribution to the thermal correction. The physical meaning of the
resulting entropic anomaly is discussed.

The paper is organized as follows. In Sec.~II, the Casimir-Polder
free energy for an atom interacting with real graphene sheet is
conveniently expressed via the polarization tensor. In Sec.~III, the
low-temperature behavior of the first contribution to the thermal
correction arising due to the Matsubara summation is found using the
polarization tensor at zero temperature. Section IV considers the
second contribution to the thermal correction arising from an
explicit temperature dependence of the polarization tensor in the
zero-frequency Matsubara term. In Sec.~V, the third contribution to
the thermal correction is found at low temperature which arises in
a similar manner from the sum of all terms with nonzero Matsubara
frequencies. In Sec.~VI, the
reader will find our conclusions and a discussion.
Appendixes A and B contain some details of the used asymptotic
expansions.

\section{The Casimir-Polder free energy for an atom interacting
with real graphene sheet described by the polarization tensor}

The free energy of an atom spaced at a distance $a$ from real
graphene sheet kept at temperature $T$ in thermal equilibrium with
the environment has the form following from the Lifshitz theory
for an atom interacting with any plate or some planar structure \cite{31}.
For our purposes, it is convenient to present this equation in terms
of dimensionless Matsubara frequencies $\zeta_l=\xi_l/\omega_c$, where
$l=0,\,1,\,2,\,\ldots$, $\xi_l=2\pi k_BTl/\hbar$ are the standard
dimensional Matsubara frequencies, and $\omega_c=c/(2a)$ is the
characteristic frequency. We also use the dimensionless integration
variable $y$ which is connected with the magnitude of the wave vector
projection on the plane of a plate $k_{\bot}$ by
$y=2a(k_{\bot}^2+\xi_l^2/c^2)^{1/2}$. Then the Casimir-Polder free
energy is expressed as
\begin{eqnarray}
&&
\cF=-\frac{k_BT}{8a^3}\sum_{l=0}^{\infty}
{\vphantom{\sum}}^{\prime}\alpha_l
\int_{\zeta_l}^{\infty}dye^{-y}
\label{eq1} \\
&&~\times
\left[(2y^2-\zeta_l^2)\rM\zy-\zeta_l^2\rE\zy\right],
\nonumber
\end{eqnarray}
\noindent
where $\alpha_l\equiv\alpha(\ri\omega_c\zeta_l)$ is the atomic electric
polarizability, the prime on the summation sign divides the term with
$l=0$ by two, and $\rM$, $\rE$ are the reflection coefficients of
electromagnetic waves with the transverse magnetic (TM) and transverse
electric (TE) polarizations on the plate (planar structure).

For a dielectric plate described by some phenomenological dielectric
permittivity, $\rM$ are $\rE$ are the standard Fresnel reflection
coefficients. However, for graphene the reflection coefficients are expressed
via the polarization tensor of graphene found on the basis of first
principles of thermal quantum field theory \cite{67,68,69,70}.
For us it is convenient to use the dimensionless polarization tensor
$\tp_{mn,l}\yt\equiv\tp_{mn}(\ri\zeta_l,y,T,\Delta,\mu)$, where
$m,\,n=0,\,1,\,2$ are the tensor indices and $l$ is the index of the
Matsubara summation defined above. The tensor $\tp_{mn,l}$
 is expressed via the dimensional one by $\tp_{mn,l}=2a\Pi_{mn,l}/\hbar$.
Then the reflection coefficients on a graphene sheet take the form
\cite{67,68,69,70}
\begin{eqnarray}
&&
\rM\zy=\frac{y\tp_{00,l}\yt}{y\tp_{00,l}\yt+2(y^2-\zeta_l^2)},
\nonumber \\[-1mm]
&&\label{eq2}\\[-2mm]
&&
\rE\zy=-\frac{\tp_{l}\yt}{\tp_{l}\yt+2y(y^2-\zeta_l^2)}.
\nonumber
\end{eqnarray}
\noindent
Here, the quantity $\tp_{l}\yt\equiv\tp(\ri\zeta_l,y,T,\Delta,\mu)$ is not
a tensor, but the following linear combination of the trace of the polarization
tensor $\tp_m^{\,m}(\ri\zeta_l,y,T,\Delta,\mu)$ and its 00 component
$\tp_{00}(\ri\zeta_l,y,T,\Delta,\mu)$:
\begin{eqnarray}
&&
\tp(\ri\zeta_l,y,T,\Delta,\mu)=
(y^2-\zeta_l^2)\tp_m^{\,m}(\ri\zeta_l,y,T,\Delta,\mu)
\nonumber \\
&&~~~~~~~~~~~
-y^2\tp_{00}(\ri\zeta_l,y,T,\Delta,\mu).
\label{eq3}
\end{eqnarray}
\noindent
(Note that in the literature this combination is usually notated by the
same letter as the tensor components which does not
create a confusion because it does not have the tensor indices.)
It was shown \cite{68} that the polarization tensor of graphene is completely
determined by its 00 component and by its trace, but in numerous applications
it is more convenient \cite{69} to use, in addition to $\tp_{00,l}$,
not the trace itself but the quantity $\tp_l$ defined in Eq.~(\ref{eq3}).

For real graphene sheet the quantities $\tp_{00,l}$ and $\tp_l$ depend on
the energy gap $\Delta$ and chemical potential $\mu$.
(Thus, the reflection coefficients also depend on $\Delta$ and $\mu$,
but we do not explicitly indicate this dependence for the sake of
brevity.) Note that a nonzero energy gap in the spectrum of electronic excitations
arises under the influence of electron-electron interaction, defects of the
crystal structure, for graphene deposited on a substrate etc. \cite{56,75,76},
whereas the value of the chemical potential is connected with the doping
concentration \cite{77}.
Explicit expressions for  $\tp_{00,l}$ and $\tp_l$ can be conveniently
presented as the sums of independent and dependent on $\mu$ and $T$ parts \cite{66}
\begin{eqnarray}
&&
\tp_{00,l}\yt= \tp_{00,l}^{(0)}(y,\Delta)+ \tp_{00,l}^{(1)}\yt,
\nonumber \\[-1mm]
&&
\label{eq4}\\[-2mm]
&&
\tp_{l}\yt= \tp_{l}^{(0)}(y,\Delta)+ \tp_{l}^{(1)}\yt.
\nonumber
\end{eqnarray}

As the independent on $\mu$ and $T$ parts on the right-hand side of Eq.~(\ref{eq4})
we take the respective quantities from Refs.~\cite{66,67}:
\begin{eqnarray}
&&
\tp_{00,l}^{(0)}(y,\Delta)=\alpha\frac{y^2-\zeta_l^2}{p_l}
\Psi\left(\frac{D}{p_l}\right),
\nonumber \\[-1mm]
&&
\label{eq5}\\[-2mm]
&&
\tp_{l}^{(0)}(y,\Delta)=\alpha(y^2-\zeta_l^2){p_l}
\Psi\left(\frac{D}{p_l}\right).
\nonumber
\end{eqnarray}
\noindent
Here, $D\equiv\Delta/(\ho)$, the function $\Psi(x)$ is defined as
\begin{equation}
\Psi(x)=2\left[x+(1-x^2)\arctan(x^{-1})\right],
\label{eq6}
\end{equation}
\noindent
$\alpha=e^2/(\hbar c)$ [in SI units $e^2/(4\pi\epsilon_0\hbar c)$
where $\epsilon_0$ is the permittivity of the vacuum]
and $\vF=v_F/c\approx 1/300$ are the fine structure
constant and the Fermi velocity normalized to the speed of light, and
\begin{equation}
p_l=\left[\vF^2y^2+(1-\vF^2)\zeta_l^2\right]^{1/2}.
\label{eq7}
\end{equation}

The $\mu$-dependent  parts on the right-hand side of Eq.~(\ref{eq4})
are more complicated. They are given by Eqs.~(13) and (14) in Ref.~\cite{66}
where it is convenient to replace the integration variable $u$ with
$t=\hbar c p_lu/(2a\Delta)$
\begin{eqnarray}
&&
\tp_{00,l}^{(1)}\yt=\frac{4\alpha D}{\vF^2}\int_1^{\infty}\!\!\!dt
w(t,T,\Delta,\mu)X_{00,l}(t,y,D),
\nonumber\\[-1mm]
&&\label{eq8}\\[-2mm]
&&
\tp_{l}^{(1)}\yt=-\frac{4\alpha D}{\vF^2}\int_1^{\infty}\!\!\!dt
w(t,T,\Delta,\mu)X_{l}(t,y,D),
\nonumber
\end{eqnarray}
\noindent
where $w$ is defined as
\begin{equation}
w(t,T,\Delta,\mu)=\left(e^{\frac{t\Delta+2\mu}{2k_BT}}+1\right)^{-1}+
\left(e^{\frac{t\Delta-2\mu}{2k_BT}}+1\right)^{-1}
\label{eq9}
\end{equation}
\noindent
and the quantities $X_{00,l}$ and $X_l$ are given by
\begin{widetext}
\begin{eqnarray}
&&
X_{00,l}(t,y,D)=1-{\rm Re}
\frac{p_l^2-D^2t^2+2\ri\zeta_lDt}{\left[p_l^4-p_l^2D^2t^2+\vF^2(y^2-\zeta_l^2)D^2
+2\ri\zeta_lp_l^2Dt\right]^{1/2}},
\nonumber\\[-1mm]
&&\label{eq10}\\[-2mm]
&&
X_{l}(t,y,D)=\zeta_l^2-{\rm Re}
\frac{\zeta_l^2p_l^2-p_l^2D^2t^2+\vF^2(y^2-\zeta_l^2)D^2+
2\ri\zeta_lp_l^2Dt}{\left[p_l^4-p_l^2D^2t^2+\vF^2(y^2-\zeta_l^2)D^2
+2\ri\zeta_lp_l^2Dt\right]^{1/2}}.
\nonumber
\end{eqnarray}
\end{widetext}

As noted in Sec.~I, we are interested to investigate the thermal
correction to the Casimir-Polder energy as a function of temperature.
For this purpose the Casimir-Polder free energy is presented in the form
\begin{equation}
\cF=E(a)+\dT\cF,
\label{eq11}
\end{equation}
\noindent
where the Casimir-Polder energy is given by
\begin{eqnarray}
&&
E(a)=-\frac{\hbar c}{32\pi a^4}\int_{0}^{\infty}\!\!\!
d\zeta\alpha(\ri\omega_c\zeta)
\int_{\zeta}^{\infty}\!\!\!dye^{-y}
\label{eq12} \\
&&~\times
\left[(2y^2-\zeta^2)\rM(\ri\zeta,y,0)-\zeta^2\rE(\ri\zeta,y,0\right],
\nonumber
\end{eqnarray}
\noindent
and the thermal correction vanishes with vanishing temperature
\begin{equation}
\lim_{T\to 0}\dT\cF=0.
\label{eq13}
\end{equation}
\noindent
The reflection coefficients in Eq.~(\ref{eq12}) are given by Eq.~(\ref{eq2})
taken at $T=0$. They are expressed via the polarization tensor of graphene
calculated at zero temperature and contain a continuous parameter $\zeta$
in place of the discrete Matsubara frequencies $\zeta_l$.

An important point is that in the limiting case of zero chemical potential,
$\mu\to 0$, the quantities $\tp_{00,l}^{(0)}$ and $\tp_{l}^{(0)}$ defined
in Eqs.~(\ref{eq4}) and (\ref{eq5}) just have the meaning of the 00 component
of the polarization tensor at zero temperature and the combination of its
components defined in Eq.~(\ref{eq3}):
\begin{eqnarray}
&&
\tp_{00,l}^{(0)}(y,\Delta)=\tp_{00,l}(y,0,\Delta,0),
\nonumber \\
&&
\tp_{l}^{(0)}(y,\Delta)=\tp_{l}(y,0,\Delta,0).
\label{eq14}
\end{eqnarray}
\noindent
In this case the quantities $\tp_{00,l}^{(1)}$ and $\tp_{l}^{(1)}$ have
the meaning of the thermal corrections to the zero-temperature
polarization tensor:
\begin{eqnarray}
&&
\tp_{00,l}^{(1)}(y,T,\Delta,0)=\dT\tp_{00,l}(y,T,\Delta,0),
\nonumber \\
&&
\tp_{l}^{(1)}(y,T,\Delta,0)=\dT\tp_{l}(y,T,\Delta,0),
\label{eq15}
\end{eqnarray}
\noindent
which goes to zero with vanishing $T$.

According to results of Ref.~\cite{66}, similar situation holds for
$\mu\neq 0$ satisfying the condition $\Delta>2\mu$. Under this condition
the polarization tensor at zero temperature does not depend on $\mu$,
so that, once again, we have
\begin{eqnarray}
&&
\tp_{00,l}^{(0)}(y,\Delta)=\tp_{00,l}(y,0,\Delta,\mu),
\nonumber \\
&&
\tp_{l}^{(0)}(y,\Delta)=\tp_{l}(y,0,\Delta,\mu),
\label{eq16}
\end{eqnarray}
\noindent
and
\begin{eqnarray}
&&
\tp_{00,l}^{(1)}\yt=\dT\tp_{00,l}\yt,
\nonumber \\
&&
\tp_{l}^{(1)}\yt=\dT\tp_{l}\yt,
\label{eq17}
\end{eqnarray}
\noindent
where the quantities in Eq.~(\ref{eq17}) go to zero when $T$ goes to zero.

Another situation takes place for $\mu\neq 0$ satisfying the condition
$\Delta<2\mu$. In this case the quantities $\tp_{00,l}^{(0)}$ and $\tp_{l}^{(0)}$
are not equal to the 00 component
of the polarization tensor at zero temperature and to the combination of its
components defined in Eq.~(\ref{eq3}).
In fact under the condition $\Delta<2\mu$ the polarization tensor at $T=0$
depends on $\mu$. The precise expressions for the quantities
$\tp_{00,l}(y,0,\Delta,\mu)$ and $\tp_{l}(y,0,\Delta,\mu)$ in this case
have been obtained in Eqs.~(21) and (24) of Ref.~\cite{78} by direct calculation
using Eqs.~(\ref{eq4})--(\ref{eq10}). In terms of the dimensionless variables
used above they are given by
\begin{widetext}
\begin{eqnarray}
&&
\tp_{00,l}(y,0,\Delta,\mu)=\frac{8\alpha\mu}{\vF^2\ho}-
\frac{2\alpha(y^2-\zeta_l^2)}{p_l^3}\left\{
\vphantom{\left[\frac{\pi}{2}\right]}(p_l^2+D^2)
{\rm Im}\left(z_l\sqrt{1+z_l^2}\right)\right.
\nonumber \\
&&~~~~~~~\left.
+(p_l^2-D^2)\left[
{\rm Im}\ln\left(z_l+\sqrt{1+z_l^2}\right)-\frac{\pi}{2}\right]\right\},
\nonumber \\[-1mm]
&&\label{eq18}\\[-1mm]
&&
\tp_{l}(y,0,\Delta,\mu)=-\frac{8\alpha\mu\zeta_l^2}{\vF^2\ho}+
\frac{2\alpha(y^2-\zeta_l^2)}{p_l}\left\{
\vphantom{\left[\frac{\pi}{2}\right]}(p_l^2+D^2)
{\rm Im}\left(z_l\sqrt{1+z_l^2}\right)\right.
\nonumber \\
&&~~~~~~~\left.
-(p_l^2-D^2)\left[
{\rm Im}\ln\left(z_l+\sqrt{1+z_l^2}\right)-\frac{\pi}{2}\right]\right\},
\nonumber
\end{eqnarray}
\end{widetext}
where
\begin{equation}
z_l\equiv z_l(y,\Delta,\mu)=\frac{p_l}{\vF\sqrt{p_l^2+D^2}\sqrt{y^2-\zeta_l^2}}
\left(\zeta_l+\ri\frac{2\mu}{\ho}\right).
\label{eq19}
\end{equation}
\noindent
It is easily seen that for $\mu=\Delta=0$ these equations reduce to the
result given by Eq.~(\ref{eq5}) with $\Delta=0$.

Now we are in a position to present the reflection coefficients (\ref{eq2})
in the form
\begin{equation}
r_{\rm TM(TE)}\zy=r_{\rm TM(TE)}\ozy+\dT r_{\rm TM(TE)}\zy,
\label{eq20}
\end{equation}
\noindent
where the first contributions on the right-hand side are determined
by the polarization tensor at $T=0$
\begin{eqnarray}
&&
\rM\ozy=\frac{y\tp_{00,l}\yo}{y\tp_{00,l}\yo+2(y^2-\zeta_l^2)},
\nonumber \\[-1mm]
&&\label{eq21}\\[-2mm]
&&
\rE\ozy=-\frac{\tp_{l}\yo}{\tp_{l}\yo+2y(y^2-\zeta_l^2)},
\nonumber
\end{eqnarray}
\noindent
whereas the second contribution has the meaning of the thermal correction
and goes to zero with vanishing $T$.
This equation, however, is valid in both cases $\Delta>2\mu$ [here,
in accordance to Eq.~(\ref{eq16}), the polarization tensor at $T=0$ is
presented in Eq.~(\ref{eq5})] and $\Delta<2\mu$ [here it is given by
Eq.~(\ref{eq18})]. As to the case $\Delta=2\mu$, it is discussed in the
next sections, as well as the explicit approximate expressions for
thermal corrections to the reflection coefficients on the right-hand side
of Eq.~(\ref{eq20}).

Using
Eq.~(\ref{eq11}), we present the thermal correction to the Casimir-Polder
energy as
\begin{equation}
\dT\cF=\cF-E(a).
\label{eq22}
\end{equation}
\noindent
Now we substitute Eq.~(\ref{eq20}) in the expression (\ref{eq1}) for
the Casimir-Polder free energy and identically present the thermal
correction $\dT\ocF$ as a sum of three contributions
\begin{equation}
\dT\cF=\daT\cF+\dbT\cF+\dcT\cF.
\label{eq22aa}
\end{equation}
\noindent
Here, the following notations are introduced:
\begin{eqnarray}
&&
\daT\cF\equiv -\frac{k_BT}{8a^3}\sum_{l=0}^{\infty}
{\vphantom{\sum}}^{\prime}\alpha_l
\int_{\zeta_l}^{\infty}dye^{-y}
\nonumber \\
&&~\times
\left[(2y^2-\zeta_l^2)\rM\ozy-\zeta_l^2\rE\ozy\right]-E(a),
\label{eq22a}
\end{eqnarray}
\noindent
where $E(a)$ is defined in Eq.~(\ref{eq12}), and
\begin{eqnarray}
&&
\dbT\cF+\dcT\cF\equiv -\frac{k_BT}{8a^3}\sum_{l=0}^{\infty}
{\vphantom{\sum}}^{\prime}\alpha_l
\int_{\zeta_l}^{\infty}\!\!\!dye^{-y}
\nonumber \\
&&~\times
\left[(2y^2-\zeta_l^2)\dT\rM\zy-\zeta_l^2\dT\rE\zy\right],
\label{eq22b}
\end{eqnarray}
\noindent
where $\dbT\ocF$ and $\dcT\ocF$ are equal to the term with $l=0$ and
to the sum of the terms with $l\geqslant 1$, respectively, in Eq.~(\ref{eq22b}).

{}From Eq.~(\ref{eq22a}) it is seen that the contribution to the thermal correction $\daT\ocF$
contains only the reflection coefficients at zero temperature. Thus, its
temperature dependence is
completely determined by a summation over the Matsubara frequencies.
For this reason, it is called ``implicit". As to the contributions $\dbT\ocF$ and $\dcT\ocF$ to the
thermal correction, defined in Eq.~(\ref{eq22b}),
they depend on the thermal corrections to the reflection coefficients
which vanish if the polarization tensor does not depend on
temperature as a parameter. Because of this, the contributions $\dbT\ocF$
and $\dcT\ocF$ are called ``explicit". In fact the quantity
in Eq.~(\ref{eq22b}) could be considered as one explicit contribution
to the thermal correction.
However, the asymptotic behaviors of $\dbT\ocF$ and $\dcT\ocF$ with
vanishing $T$ are not similar (see Secs.~IV and V), and this fact warrants a
division of this contribution into two parts.

In the next sections, the low-temperature behaviors of the thermal corrections
$\daT\ocF$, $\dbT\ocF$ and $\dcT\ocF$ are investigated
one after another.

\section{Thermal correction to the Casimir-Polder energy due to
Matsubara summation using the zero-temperature reflection
coefficients}

In this section, we find the behavior of the first contribution to the
thermal correction, $\daT\ocF$, at low temperature under different
relationships between the energy gap and chemical potential.
As defined in Eq.~(\ref{eq22a}), $\daT\ocF$ is given by the difference of the
sum over the discrete Matsubara frequencies with the reflection
coefficients  $r_{\rm TM(TE)}\ozy$, and
the integral  with respect to continuous imaginary
frequency containing the reflection coefficients $r_{\rm TM(TE)}(\ri\zeta,y,0)$.
Using the Abel-Plana formula, this difference can be written in the form
\cite{31,72a}
\begin{equation}
\daT\cF=-\ri\frac{\alpha_0k_BT}{8a^3}\int_{0}^{\infty}\!\!dt
\frac{\Phi(\ri t\tau)-\Phi(-\ri t\tau)}{e^{2\pi t}-1},
\label{eq23}
\end{equation}
\noindent
where $\Phi(x)=\Phi_1(x)+\Phi_2(x)$,
\begin{eqnarray}
&&
\Phi_1(x)=2\int_x^{\infty}\!\!dy y^2e^{-y}\rM(\ri x,y,0),
\label{eq24} \\
&&
\Phi_2(x)=-x^2\int_x^{\infty}\!\!dy e^{-y}[\rM(\ri x,y,0)+
\rE(\ri x,y,0)]
\nonumber
\end{eqnarray}
\noindent
and the dimensionless temperature parameter is defined as
$\tau=4\pi ak_BT/(\hbar c)=2\pi k_BT/(\ho)$.

In Eq.~(\ref{eq23}), we have preserved only the static atomic polarizability
$\alpha_0=\alpha(0)$ in the expansion of $\alpha(\ri t\tau)=\alpha(\ri x)$
in the powers of $x$.
This is because we are looking for the main term in the expansion of the
Casimir-Polder free energy $\ocF$ at low $T$ ($\tau\ll 1$).
Note also that care must be exercised when expanding the functions
$\Phi_1$ and $\Phi_2$ in the powers of $x$. It may happen that an expansion
of the reflection coefficients in the powers of $x$ with subsequent integration
leads to incorrect results because common powers of $x$  arise from different
expansion orders of the reflection coefficients (see below).

We begin with the case of a slightly doped graphene $\Delta>2\mu$.
In this case the polarization tensor at zero temperature does not depend
on $\mu$ [see Eq.~(\ref{eq16})] and is given by Eq.~(\ref{eq5}).
The reflection coefficients entering the thermal correction $\daT\cF$
are given by Eq.~(\ref{eq21}). For the function $\Phi_1$, defined in
Eq.~(\ref{eq24}), it is not productive to expand the reflection coefficient
$\rM$ in powers of $x$ with subsequent integration as noted above.
Instead, an expansion of $\Phi_1$ in the Taylor series in powers of $x$
using Eqs.~(\ref{eq5}) and (\ref{eq21}) results in
$\Phi_1^{\prime}(0)=\Phi_1^{(3)}(0)=\Phi_1^{(5)}(0)=0$.
Then we conclude that the leading contribution of $\Phi_1$ to $\daT\cF$
is of higher order than $T^6$ because the even powers in $x$ do not contribute
to Eq.~(\ref{eq23}).

An expansion of the function $\Phi_2$ defined in Eq.~(\ref{eq24}) in powers
of $x$ can be found by expanding the sum of the reflection coefficients
$\rM$ and $\rE$ in powers of $x$ with subsequent integration with respect
to $y$. This is done under an assumption $D>1$ which is valid at sufficiently
large separations $a>1~\mu$m. Taking into account that the main contributions
to the integrals in Eq.~(\ref{eq24}) are given by $y\sim 1$ and that
$\zeta_l=\tau l$, at low temperature the quantity $p_l$ defined in Eq.~(\ref{eq7})
satisfies the inequality $p_l\ll 1$, so that $D/p_l\gg 1$.
Then the main contribution to the function $\Psi$ in Eqs.~(\ref{eq5}) and
(\ref{eq6}) is given by
\begin{equation}
\Psi\left(\frac{D}{p_l}\right)\approx \frac{8}{3}\,\frac{p_l}{D}.
\label{eq25}
\end{equation}
\noindent
With account of this equation one obtains \cite{74}
\begin{equation}
\Phi_2(x)=\frac{\hbar c\alpha(1+\vF^2)}{3\vF^2a\Delta}x^4{\rm Ei}(-x)+Cx^4+O(x^5),
\label{eq26}
\end{equation}
\noindent
where ${\rm Ei}(z)$ is the exponential integral and $C$ is a constant which does
not contribute to Eq.~(\ref{eq23}).

Substituting Eq.~(\ref{eq26}) in Eq.~(\ref{eq23}), one finds \cite{74}
\begin{equation}
\daT\cF=-\frac{\alpha_0(k_BT)^5}{(\hbar c)^3\Delta}\,
\frac{8\alpha(1+\vF^2)}{\vF^2}.
\label{eq27}
\end{equation}
\noindent
Thus, under a condition $\Delta>2\mu$ the thermal correction $\daT\ocF$
vanishes with temperature faster than for a pristine graphene where it is
of the order of $(k_BT)^4$ \cite{72a}.

We are coming now to the case of $\Delta<2\mu$. In this case the thermal
correction $\daT\ocF$ is again given by the difference of the sum
Eq.~(\ref{eq1}) with the zero-temperature reflection coefficients
$r_{\rm TM(TE)}\ozy$ and the integral (\ref{eq12}) resulting in Eq.~(\ref{eq23}).
It is convenient, however, to present the function $\Phi(x)$ in an equivalent
form $\Phi(x)=\chi_1(x)+\chi_2(x)$, where
\begin{eqnarray}
&&
\chi_1(x)=\int_x^{\infty}\!\!dy e^{-y}(2y^2-x^2)\rM(\ri x,y,0),
\nonumber\\
&&
\chi_2(x)=-x^2\int_x^{\infty}\!\!dy e^{-y}\rE(\ri x,y,0).
\label{eq28}
\end{eqnarray}

The reflection coefficients are again given by Eq.~(\ref{eq21}), but the
polarization tensor is now presented in Eqs.~(\ref{eq18}) and (\ref{eq19})
where the discrete Matsubara frequencies $\zeta_l=\tau l$ are replaced with $x$.
Then the polarization tensor in Eq.~(\ref{eq21}) is replaced with
$\tp_{00}(x,y,0,\Delta,\mu)$. The low-temperature expansion of the quantity
$\chi_1(x)$ can be performed in the same way as of $\Phi_1(x)$, i.e., by
expanding $\chi_1(x)$ in the Taylor series in powers of $x$.
Using Eqs.~(\ref{eq28}) and (\ref{eq21}), one obtains
%\begin{widetext}
\begin{eqnarray}
&&
\chi_1^{\prime}(0)=2\int_0^{\infty}\!\!dye^{-y}y^2
\left.\frac{\partial}{\partial x}
\frac{y\tp_{00}(x,y,0,\Delta,\mu)}{y\tp_{00}(x,y,0,\Delta,\mu)+2(y^2-x^2)}
\right\vert_{x=0}
\nonumber \\
&&~~~~~~~
=4\int_{0}^{\infty}\!\!dye^{-y}y^3
\frac{\left. \frac{\partial}{\partial x}\tp_{00}(x,y,0,\Delta,\mu)
\right\vert_{x=0}}{[\tp_{00}(0,y,0,\Delta,\mu)+2y]^2}.
\label{eq29}
\end{eqnarray}
%\end{widetext}

In what follows we use the condition
\begin{equation}
\sqrt{4\mu^2-\Delta^2}>\ho,
\label{eq30}
\end{equation}
\noindent
which is valid at sufficiently large separations. Then from the first formula
in Eq.~(\ref{eq18}) we have
\begin{equation}
\tp_{00}(0,y,0,\Delta,\mu)=\frac{8\alpha}{\vF^2}\,\frac{\mu}{\ho}\equiv Q_0.
\label{eq31}
\end{equation}

By calculating the derivative of the first formula in Eq.~(\ref{eq18}) at
$x=0$, we obtain
\begin{equation}
{\left. \frac{\partial}{\partial x}\tp_{00}(x,y,0,\Delta,\mu)
\right\vert_{x=0}}=-\frac{4\alpha}{\vF^3y}\,
\frac{4\mu^2-(\ho\vF y)^2}{\sqrt{4\mu^2-(\ho\vF y)^2-\Delta^2}}.
\label{eq32}
\end{equation}
\noindent
Taking into account the condition (\ref{eq30}), the inequality $\vF\ll 1$ and
the fact that the main contribution to Eq.~(\ref{eq29}) is given by $y\sim 1$,
Eq.~(\ref{eq32}) can be simplified to
\begin{equation}
\left.\frac{\partial}{\partial x}\tp_{00}(x,y,0,\Delta,\mu)
\right\vert_{x=0}=-\frac{16\alpha}{\vF^3y}\,
\frac{\mu^2}{\ho\sqrt{4\mu^2-\Delta^2}}.
\label{eq33}
\end{equation}

Substituting Eqs.~(\ref{eq31}) and (\ref{eq33}) in Eq.~(\ref{eq29}), one finds
\begin{widetext}
\begin{eqnarray}
&&
\chi_1^{\prime}(0)=-\frac{16\alpha\mu^2}{\vF^3\ho\sqrt{4\mu^2-\Delta^2}}
\int_0^{\infty}\!\!\!dye^{-y}\frac{4y^2}{(2y+Q_0)^2}
\nonumber \\
&&~~
=-\frac{16\alpha\mu^2}{\vF^3\ho\sqrt{4\mu^2-\Delta^2}}
\left[\frac{2+Q_0}{2}+\frac{Q_0(Q_0+4)}{4}e^{Q_0/2}
{\rm Ei}\left(-\frac{Q_0}{2}\right)\right].
\label{eq34}
\end{eqnarray}
\end{widetext}

Now we have the desired result
\begin{equation}
\chi_1(x)=\chi_1(0)+\chi_1^{\prime}(0)x+O(x^2),
\label{eq36}
\end{equation}
\noindent
where $\chi_1(0)$ does not contribute to Eq.~(\ref{eq23}) and the value
of the first derivative at $x=0$ is presented in Eq.~(\ref{eq34}).
From Eq.~(\ref{eq28}) it is easily seen that $\chi_2(0)=\chi_2^{\prime}(0)=0$
and similar expansion for the function $\chi_2(x)$ takes the form
\begin{equation}
\chi_2(x)=Cx^2+O(x^3),
\label{eq37}
\end{equation}
\noindent
where $C$ is a constant which does not contribute to Eq.~(\ref{eq23}).
Thus, from Eq.~(\ref{eq36})
\begin{equation}
\Phi(\ri\tau t)-\Phi(-\ri\tau t)=2\ri\chi_1^{\prime}(0)\tau t
\label{eq38}
\end{equation}
\noindent
and, after substitution to Eq.~(\ref{eq23}) with account of Eq.~(\ref{eq34}),
one obtains
\begin{eqnarray}
&&
\daT\cF=-\frac{\alpha_0\mu^2(k_BT)^2}{(\hbar c)^2a\sqrt{4\mu^2-\Delta^2}}\,
\frac{16\alpha}{\vF^3}
\nonumber \\
&&~\times
\left[2+{Q_0}+\frac{Q_0(Q_0+4)}{2}e^{Q_0/2}
{\rm Ei}\left(-\frac{Q_0}{2}\right)\right].
\label{eq39}
\end{eqnarray}

It is seen that in the case $\Delta<2\mu$ the behavior of the thermal
correction $\daT\ocF$ at low temperature is different from the case
of graphene with $\Delta>2\mu$ [see Eq.~(\ref{eq27})] and from the case
of pristine graphene.

Now we consider the low-temperature behavior of $\daT\ocF$ for the
case $\Delta=2\mu$. This case cannot be considered by the limiting
transition $\Delta\to 2\mu$ from our result (\ref{eq39}) obtained for
$\Delta<2\mu$ because it was derived under the condition (\ref{eq30}).

Below we show that in the case $\Delta=2\mu$
the low-temperature behavior of $\daT\ocF$  is again given by
Eq.~(\ref{eq27}) derived in the case $\Delta>2\mu$. We start from the
polarization tensor at zero temperature (\ref{eq5}) where $\zeta_l=\tau l$
is replaced with $x$. To be specific, we consider
\begin{equation}
\tp_{00}^{(0)}(x,y,\Delta)=\alpha\frac{y^2-x^2}{p(x)}
\Psi\left(\frac{D}{p(x)}\right),
\label{eq40}
\end{equation}
\noindent
where $\Psi$ is defined in Eq.~(\ref{eq6}) and
$p(x)=[\vF^2y^2+(1-\vF^2)x^2]^{1/2}$.
Under the condition $D>1$, we consider the value of $\tp_{00}^{(0)}$ at
zero $x$
\begin{equation}
\tp_{00}^{(0)}(0,y,\Delta)=\frac{\alpha y}{\vF}
\Psi\left(\frac{D}{\vF y}\right).
\label{eq41}
\end{equation}
\noindent
Expanding this quantity in powers of the small parameter $\vF y/D$,
one arrives at
\begin{equation}
\tp_{00}^{(0)}(0,y,\Delta)=\frac{8\alpha y}{\vF}
\sum_{k=0}^{\infty}(-1)^{k}\frac{k+1}{(2k+1)(2k+3)}
\left(\frac{\vF y}{D}\right)^{2k+1}.
\label{eq42}
\end{equation}

This equation is also valid at $\Delta=2\mu$. To make sure that this is the
case, we consider the first formula in Eq.~(\ref{eq18}) expressing the
zero-temperature polarization tensor in the case $\Delta<2\mu$, replace
there $\zeta_l$ with $x$, put $\mu=\Delta/2$, $x=0$ and obtain
\begin{widetext}
\begin{equation}
\tp_{00}(0,y,0,\Delta,\Delta/2)=\frac{4\alpha D}{\vF^2}-
\frac{2\alpha }{\vF^3y}\left\{\vF yD+(\vF^2y^2-D^2)
\left[{\rm Im}\ln(\ri D+\vF y)-\frac{\pi}{2}\right]\right\}.
\label{eq43}
\end{equation}
\end{widetext}

Expanding this equation in powers of $\vF y/D$, one again obtains the
right-hand side of Eq.~(\ref{eq42}) with a conclusion that
\begin{equation}
\tp_{00}^{(0)}(0,y,\Delta)= \tp_{00}(0,y,0,\Delta,\Delta/2).
\label{eq44}
\end{equation}

In  a similar way, it is easy to show that
\begin{equation}
\tp^{(0)}(0,y,\Delta)= \tp(0,y,0,\Delta,\Delta/2)
\label{eq44a}
\end{equation}
\noindent
and also
\begin{equation}
\left.\frac{\partial\tp_{00}^{(0)}}{\partial x}\right\vert_{x=0}=
\left.\frac{\partial\tp_{00}}{\partial x}\right\vert_{x=0}=0,
\quad
\left.\frac{\partial\tp^{(0)}}{\partial x}\right\vert_{x=0}=
\left.\frac{\partial\tp}{\partial x}\right\vert_{x=0}=0.
\label{eq45}
\end{equation}

We conclude that the polarization tensor at zero temperature (\ref{eq5})
and (\ref{eq18}) is continuous at the point $\Delta=2\mu$, and the thermal
correction $\daT\ocF$ at this point is really given by Eq.~(\ref{eq27}).

For a more lively presentation of the obtained results, we include them in
Table~I as new information becomes available. The first column in this Table
specifies the relationship between the values of $\Delta$ and $2\mu$.
The columns 2, 3, and 4 contain up to an order of magnitude
asymptotic expressions at low $T$ for the contributions to the thermal
correction, $\daT\ocF$,  $\dbT\ocF$, and $\dcT\ocF$,  respectively, and
indicate the reflection coefficients from which they are obtained.
The columns 5 and 6 demonstrate the resulting behaviors of the thermal
correction to the Casimir-Polder energy and entropy, respectively,
at low temperature. At the moment the column 2 includes the results
found above in Eqs.~(\ref{eq27}), (\ref{eq38}) and again (\ref{eq27}).

\section{The role of explicit temperature dependence of reflection
coefficients: Zero-frequency contribution}

In this section we consider the low-temperature behavior of the second
contribution $\dbT\ocF$ to the thermal correction defined by the term
of Eq.~(\ref{eq22b}) with $l=0$. It is given by
\begin{equation}
\dbT\cF=-\frac{\alpha_0k_BT}{8a^3}\int_0^{\infty}\!\!\!dye^{-y}y^2
\dT\rM(0,y,T).
\label{eq46}
\end{equation}

To find $\dT\rM$, we substitute the representation for the polarization
tensor
\begin{equation}
\tp_{00,l}\yt=\tp_{00,l}\yo+\dT\tp_{00,l}\yt
\label{eq47}
\end{equation}
\noindent
in the first formula in Eq.~(\ref{eq2}) and expand the obtained expression
up to the first power in small parameter
\begin{equation}
\frac{\dT\tp_{00,l}\yt}{\tp_{00,l}\yo}.
\label{eq48}
\end{equation}

The result is
\begin{equation}
\dT\rM\zy=\frac{2y(y^2-\zeta_l^2)\dT\tp_{00,l}\yt}{[y\tp_{00,l}\yo
+2(y^2-\zeta_l^2)]^2}.
\label{eq49}
\end{equation}
\noindent
In this section we use Eq.~(\ref{eq49}) at $l=0$, but in Sec.~V
below it is used at
all $l\geqslant 1$.

We start with the case $\Delta>2\mu$ where, according to Eq.~(\ref{eq17}),
$\dT\tp_{00,0}=\tp_{00,0}^{(1)}$. The latter quantity is contained in
Eqs.~(\ref{eq8})--(\ref{eq10}) taken at $l=0$. We restrict ourselves
by only the second contribution to the right-hand side of Eq.~(\ref{eq9})
(below it is shown that the first one leads to an additional exponentially
small factor in the result). Thus, the thermal correction to the polarization
tensor has the form
\begin{eqnarray}
&&
\dT\tp_{00,0}\yt=\frac{4\alpha D}{\vF^2}\int_1^{\infty}\!\!dt
\left(\eEt+1\right)^{-1}
\nonumber\\
&&~~
\times\left\{1-{\rm Re}
\frac{\vF^2y^2-D^2t^2}{\vF y[\vF^2y^2-D^2(t^2-1)]^{1/2}}\right\}.
\label{eq50}
\end{eqnarray}

The integral of the first term on the right-hand side of this equation
is given by
\begin{eqnarray}
&&
\frac{4\alpha D}{\vF^2}\int_1^{\infty}\!\!dt
\left(\eEt+1\right)^{-1}=\frac{8\alpha}{\vF^2}\frac{k_BT}{\ho}
\ln\left(1+\deE\right)
\nonumber\\
&&~~~~~
\approx\frac{8\alpha}{\vF^2}\,\frac{k_BT}{\ho}\,\deE
\label{eq51}
\end{eqnarray}
\noindent
at $k_BT\ll\Delta-2\mu$.

As shown in Appendix A, for the integral of the second term on the right-hand
side of Eq.~(\ref{eq50}) one has
\begin{eqnarray}
&&
\frac{4\alpha D}{y\vF^3}\int_1^{f(y)}\!\!dt
\left(\eEt+1\right)^{-1}
\frac{D^2t^2-\vF^2y^2}{[\vF^2y^2-D^2(t^2-1)]^{1/2}}
\nonumber\\
&&~~
<\frac{4\alpha}{\vF^2}\sqrt{D^2+\vF^2y^2}\deE
e^{-\frac{(\ho\vF y)^2}{4k_BT\Delta}},
\label{eq52}
\end{eqnarray}
\noindent
where
\begin{equation}
f(y)=\sqrt{1+\frac{\vF^2y^2}{D^2}}.
\label{eq53}
\end{equation}

The quantity in Eq.~(\ref{eq52}) contains an additional factor exponentially
small at $T\to 0$, as compared to Eq.~(\ref{eq51}), and, thus, can be neglected.
As a result, we have
\begin{equation}
\dT\tp_{00,0}\yt=\frac{8\alpha}{\vF^2}\,\frac{k_BT}{\ho}\,\deE
\label{eq54}
\end{equation}
\noindent
Note that the first contribution on the right-hand side of Eq.~(\ref{eq9})
omitted above
would lead to an additional exponentially small factor of the order
of $e^{-2\mu/(k_BT)}$.

{}From Eqs.~(\ref{eq5}) and (\ref{eq25}) one also obtains
\begin{equation}
\tp_{00,0}\yo=\frac{8}{3}\frac{\alpha\ho}{\Delta}y^2.
\label{eq55}
\end{equation}

Substituting Eqs.~(\ref{eq54}) and (\ref{eq55}) in Eq.~(\ref{eq49}) taken at
$l=0$, we find
\begin{equation}
\dT\rM\oyt=\frac{4\alpha}{\vF^2}\,\frac{k_BT}{\ho}\,\deE
\frac{1}{y(1+qy)^2},
\label{eq56}
\end{equation}
\noindent
where $q=4\alpha\ho/(3\Delta)$.

Substituting this equation in Eq.~(\ref{eq46}) and calculating the integral,
one obtains
\begin{widetext}
\begin{eqnarray}
&&
\dbT\cF=-\alpha_0\frac{\alpha(k_BT)^2}{a^2\vF^2\hbar c}\deE
\int_0^{\infty}\!\!\!dye^{-y}\frac{y}{(1+qy)^2}
\nonumber \\
&&~~
=
\alpha_0\frac{\alpha(k_BT)^2}{a^2\vF^2\hbar c}\deE
\frac{1}{q^2}\left[1+\left(1+\frac{1}{q}\right)e^{1/q}
{\rm Ei}\left(-\frac{1}{q}\right)\right].
\label{eq57}
\end{eqnarray}
\end{widetext}
\noindent
Then under the condition $D>1$ ($\Delta>\ho$) we arrive at
\begin{eqnarray}
\dbT\cF &\approx&
-\alpha_0\frac{\alpha(k_BT)^2}{a^2\vF^2\hbar c}\deE (1-4q)
\nonumber \\
&\approx&
-\alpha_0\frac{\alpha(k_BT)^2}{a^2\vF^2\hbar c}\deE.
\label{eq58}
\end{eqnarray}
\noindent
This is quite different behavior at low $T$ than that obtained in Eq.~(\ref{eq27})
for the thermal correction $\daT\ocF$ under the condition $\Delta>2\mu$.

Now we turn to the case $\Delta<2\mu$ for the thermal correction $\dbT\ocF$.
In this case the thermal correction $\dT\tp_{00,0}$ is given by
\begin{equation}
\dT\tp_{00,0}\yt=\tp_{00,0}^{(1)}\yt-\tp_{00,0}^{(1)}\yo,
\label{eq59}
\end{equation}
\noindent
where $\tp_{00,0}^{(1)}$ is defined in Eqs.~(\ref{eq8})--(\ref{eq10}).
For $l=0$ one has
\begin{equation}
\tp_{00,0}^{(1)}\yt=\frac{4\alpha D}{\vF^2}(I_1+I_2),
\label{eq60}
\end{equation}
\noindent
where
\begin{eqnarray}
&&
I_1=\int_1^{\infty}\!\!dt\left(\eEt+1\right)^{-1},
\label{eq61} \\
&&
I_2=\frac{1}{\vF y}{\rm Re}\int_1^{\infty}\!\!dt\left(\eEt+1\right)^{-1}
\nonumber \\
&&~~~~~~~~\times
\frac{D^2t^2-\vF^2y^2}{[\vF^2y^2-D^2(t^2-1)]^{1/2}}.
\nonumber
\end{eqnarray}

Similar to the case of $\Delta>2\mu$, the first exponential term on the right-hand
side of Eq.~(\ref{eq9})  leads to
additional exponentially decreasing factors when the temperature vanishes.
For this reason, we do not consider it below. Thus, according to
Eqs.~(\ref{eq59})--(\ref{eq61}), the thermal correction to the polarization
tensor takes the form
\begin{widetext}
\begin{eqnarray}
&&
\dT\tp_{00,0}\yt=\frac{4\alpha D}{\vF^2}\left[
\int_1^{\infty}\!\!\!dt\left(\eEt+1\right)^{-1}\right.
\nonumber\\
&&~
\left.\vphantom{\int_1^{2\mu/\Delta\!\!}}
-
\int_1^{2\mu/\Delta\!\!}\!\!\!dt+
I_2-\lim_{T\to 0}I_2\right]
=\frac{4\alpha }{\vF^2}
e^{-\frac{2\mu-\Delta}{2k_BT}}\left(\frac{2k_BT}{\hbar\omega_c}-1\right)
\nonumber\\
&&~~~
\approx -\frac{4\alpha}{\vF^2}\meE
\label{eq62}
\end{eqnarray}
\end{widetext}
\noindent
at sufficiently low temperature.

Substituting Eq.~(\ref{eq62}) in Eq.~(\ref{eq49}) with $l=0$ and taking into
account that $\tp_{00,0}\yo=Q_0$, where $Q_0$ is defined in Eq.~(\ref{eq31}),
one finds
\begin{equation}
\dT\rM\oyt=-\frac{8\alpha}{\vF^2}\,\meE
\frac{y}{(Q_0+2y)^2}.
\label{eq63}
\end{equation}

Finally from Eqs.~(\ref{eq46}) and (\ref{eq63})
we obtain
\begin{eqnarray}
&&
\dbT\cF=\alpha_0\frac{\alpha k_BT}{4a^3\vF^2}\meE
\int_0^{\infty}\!\!\!dye^{-y}\frac{4y^3}{\left(2y+Q_0\right)^2}
\nonumber \\
&&
=
\alpha_0\frac{\alpha k_BT}{4a^3\vF^2}\meE
\label{eq64} \\
&&
~~\times
\left[1-\frac{Q_0}{4}(4+Q_0)-\frac{Q_0^2}{8}\left(Q_0+6\right)
e^{Q_0/2}
{\rm Ei}\left(-\frac{Q_0}{2}\right)\right].
\nonumber
\end{eqnarray}

By comparing Eqs.~({\ref{eq58}) and (\ref{eq64}), one can conclude that in the case
$\Delta<2\mu$ the thermal correction $\dbT\ocF$ at low temperatures again
decreases with $T$ exponentially fast.

Let us now consider the last case $\Delta=\mu$. Similar in Sec.~III, it can be
considered starting from the results obtained for $\Delta>2\mu$.
Now, however, the last transformation in Eq.~(\ref{eq51}) is not allowed because
$\exp[(-\Delta+2\mu)/(2k_BT)]=1$. As a result, Eq.~(\ref{eq54}) should be
replaced with
\begin{equation}
\dT\tp_{00,0}\yt=\frac{8\alpha}{\vF^2}\,\frac{k_BT}{\ho}\,\ln 2.
\label{eq65}
\end{equation}

Substituting Eqs.~(\ref{eq55}) and (\ref{eq65}) in Eq.~(\ref{eq49}), we have
\begin{equation}
\dT\rM\oyt=\frac{4\alpha\ln2}{\vF^2}\,\frac{k_BT}{\ho}
\frac{1}{y(1+qy)^2}.
\label{eq66}
\end{equation}
\noindent
Then, from Eq.~(\ref{eq46}), in place of Eq.~(\ref{eq58}) we finally obtain
\begin{equation}
\dbT\cF =
-\alpha_0\frac{\alpha\ln2(k_BT)^2}{a^2\vF^2\hbar c},
\label{eq67}
\end{equation}
\noindent
i.e., the same behavior with $T$  as was found for $\daT\ocF$
in the case $\Delta<2\mu$ [see Eq.~(\ref{eq39})].

The results presented in Eqs.~(\ref{eq58}),
(\ref{eq64}), and (\ref{eq67}) are illustrated in the column 3 of Table~I.

\section{Explicit temperature dependence of reflection
coefficients: Summation over the nonzero Matsubara frequencies}

Here we consider the low-temperature behavior of the last, third, contribution
$\dcT\ocF$ to the thermal correction in Eq.~(\ref{eq22aa}) which is
determined by an explicit dependence of the polarization tensor on $T$ in
all Matsubara terms with $l\neq 0$. In accordance to Eq.~(\ref{eq22b}),
it is given by
\begin{eqnarray}
&&
\dcT\cF=-\alpha_0\frac{k_BT}{8a^3}\sum_{l=1}^{\infty}
\int_{\zeta_l}^{\infty}\!\!\!dye^{-y}G(\zeta_l^2,y,T,\Delta,\mu),
\nonumber \\
&&
G(\zeta_l^2,y,T,\Delta,\mu)=(2y^2-\zeta_l^2)\dT\rM\zy
\nonumber \\
&&~~~~~~~~~~~~~~~~~~~~~~~~~-
\zeta_l^2\dT\rE\zy.
\label{eq68}
\end{eqnarray}
\noindent
An expression for the thermal correction $\dT\rM$ is already given in Eq.~(\ref{eq49}). To derive similar expression for $\dT\rE$, we substitute
the representation
\begin{equation}
\tp_{l}\yt=\tp_{l}\yo+\dT\tp_{l}\yt
\label{eq69}
\end{equation}
\noindent
in the second formula in Eq.~(\ref{eq2}) and expand the obtained expression
up to the first power in small parameter
\begin{equation}
\frac{\dT\tp_{l}\yt}{\tp_{l}\yt}.
\label{eq70}
\end{equation}
\noindent
The desired result is given by
\begin{equation}
\dT\rE\zy=-\frac{2y(y^2-\zeta_l^2)\dT\tp_{l}\yt}{[\tp_{l}\yo
+2y(y^2-\zeta_l^2]^2}.
\label{eq71}
\end{equation}

We start with the case $\Delta>2\mu$ where, according to Eq.~(\ref{eq17}),
$\dT\tp_{00,l}=\tp_{00,l}^{(1)}$ and $\dT\tp_{l}=\tp_{l}^{(1)}$
with $\tp_{00,l}^{(1)}$ and $\tp_{l}^{(1)}$ defined in
Eqs.~(\ref{eq8})--(\ref{eq10}).
According to Eq.~(\ref{eq16}), $\tp_{00,l}\yo=\tp_{00,l}^{(0)}$ and
$\tp_{l}\yo=\tp_{l}^{(0)}$, where the right-hand sides of these equations
are given by Eqs.~(\ref{eq5})--(\ref{eq7}), and under the condition $D>1$
the quantity $\Psi(D/p_l)$ can be replaced with $8p_l/(3D)$ [see Eq.~(\ref{eq25})].

Then, in the lowest order of the small parameter $p_l/D$, Eqs.~(\ref{eq49})
and (\ref{eq71}) take the form
\begin{eqnarray}
&&
\dT\rM\zy=\frac{2y\dT\tp_{00,l}\yt}{(y^2-\zeta_l^2)
\left(\frac{8\alpha y}{3D}+2\right)^2}
\nonumber \\
&&~~~~~~~\approx
\frac{y\dT\tp_{00,l}\yt}{2(y^2-\zeta_l^2)},
\nonumber \\
&&
\dT\rE\zy=-\frac{2y\dT\tp_{l}\yt}{(y^2-\zeta_l^2)
\left(\frac{8\alpha p_l^2}{3D}+2y\right)^2}
\nonumber \\
&&~~~~~~~\approx
-\frac{\dT\tp_{l}\yt}{2y(y^2-\zeta_l^2)}.
\label{eq72}
\end{eqnarray}
\noindent
Note that we have omitted two small terms, $8\alpha y/(3D)$ and
$8\alpha p_l^2/(3D)$, in the denominators because, similar to
Eqs.~(\ref{eq56})--(\ref{eq58}), they lead to the thermal corrections of
higher orders which can be neglected in the result.
The function $G$ in Eq.~(\ref{eq68}), which depends on $\zeta_l^2$, can be
expanded in the powers of $\zeta_l^2=(\tau l)^2$
\begin{equation}
G(\zeta_l^2,y,T,\Delta,\mu)=2y^2\dT\rM\oyt+\zeta_l^2\left.
\frac{\partial G}{\partial\zeta_l^2}\right\vert_{\zeta_l=0}
+\ldots\,.
\label{eq73}
\end{equation}

Substituting this expansion in Eq.~(\ref{eq68}), one obtains
\begin{equation}
\dcT\cF=-\alpha_0\frac{k_BT}{8a^3}(J_1+J_2),
\label{eq74}
\end{equation}
\noindent
where
\begin{eqnarray}
&&
J_1=2\sum_{l=1}^{\infty}\int_{\zeta_l}^{\infty}\!\!\!dye^{-y}y^2\dT\rM\oyt,
\label{eq75}\\
&&
J_2=\sum_{l=1}^{\infty}\zeta_l^2\int_{\zeta_l}^{\infty}\!\!\!dye^{-y}
\left.\frac{\partial G}{\partial\zeta_l^2}\right\vert_{\zeta_l=0}
+\ldots\,.
\nonumber
\end{eqnarray}

{}From Eqs.~({\ref{eq54}) and (\ref{eq72}) we find
\begin{equation}
\dT\rM\oyt=\frac{8\alpha k_BT a}{\vF^2\hbar cy}\deE.
\label{eq76}
\end{equation}
\noindent
Taking this into account, we rewrite the quantity $J_1$ in Eq.~(\ref{eq75}) as
\begin{eqnarray}
&&
J_1=\frac{16\alpha k_BT a}{\vF^2\hbar c}\deE
\sum_{l=1}^{\infty}\int_{\zeta_l}^{\infty}\!\!\!dye^{-y}y
\label{eq77}\\
&&~~
=\frac{16\alpha k_BT a}{\vF^2\hbar c}\deE\left[\frac{1}{e^{\tau}-1}+
\frac{\tau e^{\tau}}{(e^{\tau}-1)^2}\right]
\nonumber\\
&&~~\approx\frac{16\alpha k_BT a}{\vF^2\hbar c}\deE\frac{2}{\tau}
=\frac{8\alpha}{\vF^2\pi}\deE.
\nonumber
\end{eqnarray}

As shown in Appendix B, the integral $J_2$ contains the same exponentially fast
decreasing with $T$ factor and differs from Eq.~(\ref{eq77}) only by the
pre-exponent coefficient.
Because of this, using Eq.~(\ref{eq77}), we obtain from  Eq.~(\ref{eq74})
\begin{equation}
\dcT\cF\sim-\alpha_0\frac{k_BT}{a^3}\deE.
\label{eq78}
\end{equation}
\noindent
It is seen that here the factor in front of the exponent decreases slower than
in $\dbT\ocF$ [see Eq.~(\ref{eq58})].

The case $\Delta<2\mu$ can be considered in a similar manner.
Using Eqs.~(\ref{eq31}) and (\ref{eq49}), we have
\begin{equation}
\dT\rM\zy=\frac{2y(y^2-\zeta_l^2)\dT\tp_{00,l}\yt}{[yQ_0
+2(y^2-\zeta_l^2)]^2}.
\label{eq79}
\end{equation}
\noindent
Substituting here Eq.~(\ref{eq62}) and taking into account that under the
condition (\ref{eq30}) the inequality $Q_0\gg 1$ holds, one can neglect by $2y$
as compared to $Q_0$ and obtain
\begin{equation}
\dT\rM\oyt=-\frac{8\alpha}{\vF^2 Q_0^2}\meE y.
\label{eq80}
\end{equation}

Then the quantity $J_1$ defined in Eq.~(\ref{eq75}) is
\begin{eqnarray}
&&
J_1=-\frac{16\alpha}{\vF^2 Q_0^2}\meE
\sum_{l=1}^{\infty}\int_{\zeta_l}^{\infty}\!\!\!dye^{-y}y^3
\label{eq81}\\
&&~~
=-\frac{16\alpha}{\vF^2 Q_0^2}\meE
\left[\tau^3\frac{e^{\tau}(1+4e^{\tau}+e^{2\tau})}{(e^{\tau}-1)^4}\right.
\nonumber\\
&&~~\left.
+3\tau^2\frac{e^{\tau}(1+e^{\tau})}{(e^{\tau}-1)^3}
+6\tau\frac{e^{\tau}}{(e^{\tau}-1)^2}
+\frac{6}{e^{\tau}-1}\right]
\nonumber\\
&&~~\approx-\frac{16\alpha}{\vF^2 Q_0^2}\meE\frac{24}{\tau}
=-\frac{96\alpha\hbar c}{\vF^2 Q_0^2\pi ak_BT}\meE.
\nonumber
\end{eqnarray}
\noindent
Similar to Appendix B, it can be shown that the integral $J_2$ leads to
the same, up to a factor, dependence on $T$, as in Eq.~({\ref{eq81}).
 Thus, from Eq.~(\ref{eq74}) one finds
\begin{equation}
\dcT\cF\sim \alpha_0\frac{\hbar c}{a^4}\meE.
\label{eq82}
\end{equation}
\noindent
This  dependence should be compared with
that given by Eq.~(\ref{eq78}) for the case
$\Delta>2\mu$.

Now we consider the behavior of $\dcT\ocF$  at low temperature in
the case $\Delta=2\mu$. Similar to the correction $\dbT\ocF$ in
Sec.~IV, this behavior can be investigated using the results obtained
for $\Delta>2\mu$. For this purpose, we take into account that
$\Delta=2\mu$ and from the next to last transformation in
Eq.~(\ref{eq62}) obtain
\begin{equation}
\dT\tp_{00,0}\yt=\frac{8\alpha k_BT}{\vF^2\ho}\ln 2.
\label{eq83}
\end{equation}

Then from Eq.~(\ref{eq72}) we have
\begin{equation}
\dT\rM\oyt=\frac{8\alpha k_BT}{\vF^2\hbar c}\,\frac{\ln 2}{y}.
\label{eq84}
\end{equation}
\noindent
Repeating the same derivations as in the case $\Delta>2\mu$,
one arrives at
\begin{equation}
J_1=\frac{16\alpha k_BTa\ln 2}{\vF^2\hbar c}\,\frac{2}{\tau}
=\frac{8\alpha\ln 2}{\vF^2\pi}
\label{eq85}
\end{equation}
\noindent
and for the thermal correction $\dcT\ocF$ for $\Delta=2\mu$
finally finds
\begin{equation}
\dcT\cF\sim-\alpha_0\frac{k_BT}{a^3}.
\label{eq86}
\end{equation}

The results given by Eqs.~(\ref{eq78}), (\ref{eq82}), and (\ref{eq86})
are presented in the column 4 of Table I. A summary of columns 2, 3, and 4
in column 5 demonstrates the leading term in the asymptotic behavior of
the thermal correction to the Casimir-Polder energy at low $T$ for
any relationship between $\Delta$ and $2\mu$.

It is seen that Eq.~(\ref{eq86})
 differs fundamentally from the behaviors of all thermal
corrections considered above. According to the obtained results, in the
cases $\Delta>2\mu$ and $\Delta<2\mu$ the Casimir-Polder entropy
\begin{equation}
S(a,T)=-\frac{\partial\cF}{\partial T},
\label{eq87}
\end{equation}
\noindent
where the Casimir-Polder free energy $\cF$ is defined in
Eqs.(\ref{eq1}) and (\ref{eq11}),
vanishes with vanishing $T$. In the case $\Delta=2\mu$
the contribution to the entropy determined by the thermal corrections
$\daT\ocF$ and $\dbT\ocF$ vanishes with vanishing temperature
\begin{equation}
-\lim_{T\to 0}\frac{\partial}{\partial T}
\left[\daT\ocF+
\dbT\ocF\right]= 0.
\label{eq88}
\end{equation}

However, according to Eq.~(\ref{eq86}), the contribution to the entropy
determined by
the thermal correction $\dcT\ocF$ in the case $\Delta=2\mu$
gives rise to some kind of entropic anomaly
\begin{equation}
-\lim_{T\to 0}\frac{\partial}{\partial T}
\dcT\cF\neq 0.
\label{eq89}
\end{equation}
\noindent
As  a result, in the case $\Delta=2\mu$
the entropy at zero temperature is not equal to zero and depends
on the parameters of a system which means a violation of the Nernst heat theorem
(see the column 6 of Table I for the low-temperature behavior of the
Casimir-Polder entropy in different cases).
These results are
discussed in Sec.~VI in connection with similar problems of the Casimir
physics arising for metallic and dielectric materials.

\section{Conclusions and discussion}

In the foregoing, we have found the behavior of the Casimir-Polder
free energy and entropy at low temperature for a polarizable atom
interacting with real graphene sheet possessing nonzero energy gap
and chemical potential. As discussed in Sec. I, this subject is of
much fundamental interest in connection with problems arising in
Casimir physics when using the commonly accepted local models
of the dielectric response for both metallic and dielectric materials.
The distinctive feature of graphene is that its nonlocal dielectric
response, described by the polarization tensor, is found exactly on
the basis of first principles of thermal quantum field theory. At the
same time, the dielectric responses of conventional materials,
described, e.g., by the Drude or plasma models, are partially the
phenomenological ones. They are well confirmed experimentally only
for real electromagnetic fields on a mass-shell, although in the
Lifshitz theory the integration is made over all momenta both on
and off a mass-shell.

According to our results, the contribution $\daT\ocF$
to the thermal correction to the Casimir-Polder energy, originating
from a summation over the Matsubara frequencies using the zero-temperature
polarization tensor, behaves as $\sim(k_{B}T)^5$ and $\sim(k_{B}T)^2$ at low temperature under the
conditions $\Delta > 2\mu$ and $\Delta < 2\mu$, respectively.  The
contribution $\dbT\ocF$ to the Casimir-Polder energy,
which is caused by an explicit temperature dependence of the
polarization tensor in the zero-frequency Matsubara term, behaves as
\begin{equation}
\dbT\ocF\sim\left\{
\begin{array}{ll}
-(k_BT)^2\deE,&\quad \Delta>2\mu,\\
k_BT \meE,&\quad \Delta<2\mu.
\end{array}
\right.
\nonumber
\end{equation}
\noindent
In the case $\Delta = 2\mu$, one has
 $\dbT\ocF \sim -(k_{B}T)^2$.

The most interesting situation arises for the thermal correction
$\dcT\ocF$ originating from an explicit temperature
dependence of the polarization tensor in the sum of all Matsubara
terms with nonzero frequencies. As shown in this paper, a summation
over all nonzero Matsubara frequencies reduces by one the power of the
leading temperature dependence in each of the cases $\Delta > 2\mu$,
$\Delta < 2\mu$, and $\Delta = 2\mu$. As a result, one obtains that
\begin{equation}
\dcT\ocF\sim\left\{
\begin{array}{ll}
-k_BT\deE,&\quad \Delta>2\mu,\\
\meE,&\quad \Delta<2\mu,
\end{array}
\right.
\nonumber
\end{equation}
\noindent
and $\dcT\ocF \sim -k_{B}T$ for $\Delta = 2\mu$.

The above results for all three contributions to the thermal correction
combined together lead us
to a conclusion that in both cases $\Delta > 2\mu$ and $\Delta < 2\mu$
the Casimir-Polder free energy and entropy satisfy the Nernst heat
theorem. In doing so, the leading terms in the Casimir-Polder free
energy at low temperature behave as $ \sim(k_{B}T)^5$ and $~(k_{B}T)^2$
for $\Delta > 2\mu$ and $\Delta < 2\mu$, respectively.
Thus, our results do not support the statement of
Ref.~\cite{73} that ``the first order correction is quadratic over temperature
$\sim T^2$." This is true for the case $\Delta<2\mu$  but not for
$\Delta>2\mu$ where the total free energy $\mathcal{F}\sim (k_BT)^5$.
Also, if the exact equality $\Delta = 2\mu$
is valid, the Casimir-Polder free energy is linear in temperature
$\mathcal{F}\sim k_BT$.
In this case the Casimir-Polder entropy at zero temperature is equal to a
nonzero constant depending on the parameters of a system and, thus,
the Nernst heat theorem is violated. Note
for a pristine graphene  where $\mathcal{F} \sim (k_{B}T)^3$
the Nernst heat theorem is satisfied \cite{72a}.

As discussed in Sec. I, for dielectrics and metals the models of
dielectric response leading to a violation of the Nernst heat theorem
also result in contradictions between the theoretical predictions and
the experimental data for the Casimir and Casimir-Polder forces.
Up to date there is a single experiment on measuring the Casimir
interaction between a Au-coated sphere and a graphene sheet deposited
on a substrate \cite{79}, and its data are in good agreement with
theoretical results obtained using the polarization tensor of graphene
\cite{80}. In fact the values of $\Delta$ and $\mu$ for a graphene
sample used in the experiment are not known precisely so that from the
practical standpoint the equality $\Delta = 2\mu$ cannot be satisfied
exactly. For comparison purposes, the character of the real part of
conductivity of graphene as a function of frequency also changes
qualitatively depending on whether $\Delta > 2\mu$ or $\Delta < 2\mu$
\cite{81}, so that the condition $\Delta = 2\mu$ defines a singular
point.

One can conclude that with the only exception of a physically
unrealizable case $\Delta = 2\mu$ the Casimir-Polder free energy and
entropy for an atom interacting with real graphene sheet characterized
by nonzero energy gap and chemical potential satisfy the Nernst heat
theorem. This result provides further support to the assumption that
the widely known problems in Casimir physics discussed in Sec.~I may
be connected with the phenomenological character of local response
functions used for both metallic and dielectric materials.

%%%%%%%%%%%%%%%%%%%%%%%%%%%%%%%%%%%%%%
\section*{Acknowledgments}
The work of G.L.K. and V.M.M. was partially supported by the Peter
the Great Saint Petersburg Polytechnic University in the framework
of the Program ``5--100--2020". The work of V.M.M. was partially funded
by the Russian Foundation for Basic Research, Grant No. 19-02-00453 A.
His work was also partially supported by the Russian Government Program
of Competitive Growth of Kazan Federal University.
%%%%%%%%%%%%%%%%%%%%%%%%%%%%%%%%%%%%%%
%%%%%%%%%%%%%%%%%%%%%%%%%%%%%%%%%%%%%%%%
\appendix
\section{Bound for the contribution to {\boldmath$\dT\tp_{00,0}$} in the
case {\boldmath$\Delta>2\mu$}}
\setcounter{equation}{0}
\renewcommand{\theequation}{A\arabic{equation}}

In this Appendix, we consider the integral used in Eq.~(\ref{eq52}) and  for
$\Delta>2\mu$ and restrict it as follows:
\begin{widetext}
\begin{eqnarray}
&&
\int_1^{f(y)}\!\!dt
\left(\eEt+1\right)^{-1}
\frac{D^2t^2-\vF^2y^2}{[\vF^2y^2-D^2(t^2-1)]^{1/2}}
<D^2\int_1^{f(y)}\!\!dt
\left(\eEt+1\right)^{-1}
\nonumber\\
&&~~~~~~~~
\times\frac{t^2}{[\vF^2y^2-D^2(t^2-1)]^{1/2}}\equiv  I,
\label{A1}
\end{eqnarray}
\end{widetext}
\noindent
where $f(y)$ is defined in Eq.~(\ref{eq53}).

The integral on the right-hand side of Eq.~(\ref{A1}) can be integrated by
parts
\begin{eqnarray}
&&
I=-\int_1^{f(y)}\!\!t
\left(\eEt+1\right)^{-1}
d[\vF^2y^2-D^2(t^2-1)]^{1/2}
\nonumber \\
&&~
=\left(e^{\frac{\Delta-2\mu}{2k_BT}}+1\right)^{-1}\vF y
\label{A2}\\
&&~~
+\int_1^{f(y)}\!\!d\left[t
\left(\eEt+1\right)^{-1}\right]
[\vF^2y^2-D^2(t^2-1)]^{1/2}.
\nonumber
\end{eqnarray}
\noindent
The square root on the right-hand side of Eq.~(\ref{A2}) only increases if
to replace it with $\vF y$. Then Eq.~(\ref{A2}) transforms to
\begin{equation}
I<\vF yf(y)\left[e^{\frac{f(y)\Delta-2\mu}{2k_BT}}+1\right]^{-1}.
\label{A3}
\end{equation}

Now we take into account that, according to Eq.~(\ref{eq53}),
\begin{equation}
f(y)=\sqrt{1+\frac{\vF^2y^2}{D^2}}\approx 1+\frac{\vF^2y^2}{2D^2}
= 1+\frac{\vF^2y^2(\ho)^2}{2\Delta^2}
\label{A4}
\end{equation}
\noindent
and that for sufficiently low $T$ the inequality $\Delta-2\mu\gg 2k_BT$
holds. Then one can neglect by the unity in Eq.~(\ref{A3}) as compared
to the exponent and, substituting Eq.~(\ref{A4}) to its power, obtain
\begin{equation}
I<\frac{\vF y}{D}\sqrt{D^2+\vF^2y^2}\deE
e^{-\frac{(\ho\vF y)^2}{4k_BT\Delta}}.
\label{A5}
\end{equation}

Multiplying Eqs.~(\ref{A1}) and (\ref{A5}) by the factor $4\alpha D/(y\vF^3)$,
we arrive at Eq.~(\ref{eq52}).
%%%%%%%%%%%%%%%%%%%%%%%%%%%%%%%%%%%%%%%%
%%__Appendix__B__%%%%
\section{Estimation for the contribution to {\boldmath$\dcT\ocF$} in the
case {\boldmath$\Delta>2\mu$}}
\setcounter{equation}{0}
\renewcommand{\theequation}{B\arabic{equation}}

Here, we estimate the contribution $J_2$ to the thermal correction to
the Casimir-Polder free energy (\ref{eq74}) defined by the second expression
in Eq.~(\ref{eq75}).

According to Eq.~(\ref{eq68}), the value of the first derivative of $G$,
entering Eq.~(\ref{eq75}), is given by
\begin{eqnarray}
&&
\left.\frac{\partial G}{\partial\zeta_l^2}\right\vert_{\zeta_l=0}=
-\dT\rM\oyt
\label{B1} \\
&&~~~
+2y^2\left.\frac{\partial }{\partial\zeta_l^2}
\dT\rM\zy\right\vert_{\zeta_l=0}
-\dT\rE\oyt,
\nonumber
\end{eqnarray}
where expressions for the thermal corrections to the reflection coefficients
are contained in Eq.~(\ref{eq72}).

The derivative of the thermal correction $\dT\rM$ is calculated using the
first expression in Eq.~(\ref{eq72})
\begin{eqnarray}
&&
\left.\frac{\partial }{\partial\zeta_l^2}
\dT\rM\zy\right\vert_{\zeta_l=0}=\frac{1}{2y^3}\dT\tp_{00,0}\yt
\nonumber \\
&&~~~~~~
+\frac{1}{2y}
\left.\frac{\partial }{\partial\zeta_l^2}
\dT\tp_{00,l}\yt\right\vert_{\zeta_l=0}
.
\label{B2}
\end{eqnarray}

Using Eq.~(\ref{eq17}) and Eqs.~(\ref{eq8})--(\ref{eq10}), where only the
second term contributes in Eq.~(\ref{eq9}), one obtains at sufficiently
low $T$
\begin{eqnarray}
&&
\left.\frac{\partial }{\partial\zeta_l^2}
\dT\tp_{00,l}\yt\right\vert_{\zeta_l=0}=\frac{4\alpha D}{\vF^2}
\nonumber \\
&&~~~~~~~
\times
\int_1^{\infty}\!\!\!dt\left(\eEt+1\right)^{-1}
\left.\frac{\partial \chi_{00,l}}{\partial\zeta_l^2}\right\vert_{\zeta_l=0}
\nonumber \\
&&~~~~~~~~~~~~
\approx\frac{b_1}{y^4}\,\frac{k_BT}{\ho}\deE,
\label{B3}
\end{eqnarray}
\noindent
where the numerical value of the constant $b_1$ is of no concern for us now.

As to the thermal correction $\dT\tp_{00.0}$ in Eq.~(\ref{B2}), at low $T$
it is contained in Eq.~(\ref{eq54}) and can be written in the form
\begin{equation}
\dT\tp_{00,0}\yt=b_2\frac{k_BT}{\ho}\deE.
\label{B4}
\end{equation}
\noindent
Substituting Eqs.~(\ref{B3}) and (\ref{B4}) in Eq.~(\ref{B2}), one obtains
\begin{equation}
\left.\frac{\partial }{\partial\zeta_l^2}
\dT\rM\zy\right\vert_{\zeta_l=0}=\frac{k_BT}{2\ho}\left(
\frac{b_1}{y^5}+\frac{b_2}{y^3}\right)\deE.
\label{B5}
\end{equation}

Now we return to Eq.~(\ref{B1}) where the first term on the right-hand side is
found from Eqs.~(\ref{eq72}) and (\ref{B4})
\begin{equation}
\dT\rM\oyt=\frac{b_2}{2y}\frac{k_BT}{\ho}\deE.
\label{B6}
\end{equation}

The remaining term $\dT\rE\oyt$ in Eq.~(\ref{B1}) is given by the second
expression in Eq.~(\ref{eq72}) where, in accordance to Eq.~(\ref{eq17}),
the quantity $\dT\tp_{0}$ is expressed by Eq.~(\ref{eq5}).
The following result is found with the help of Eqs.~(\ref{eq8})--(\ref{eq10}):
\begin{eqnarray}
&&
\dT\rE\oyt=\frac{4\alpha D^3y}{\vF}\int_1^{f(y)}\!\!\!dt
\left(\eEt+1\right)^{-1}
\nonumber \\
&&~~~~~~~\times
\frac{t^2-1}{[\vF^2y^2-D^2(t^2-1)]^{1/2}},
\label{B7}
\end{eqnarray}
\noindent
where $f(y)$ is defined in Eq.~(\ref{eq53}). This quantity is similar to that
considered in Appendix A. By repeating the derivations of Appendix A, it is
easy to see that it contains an exponentially decreasing with $T$ factor in
addition to that one contained in Eq.~(\ref{B6}). Thus, we can neglect  by the
quantity (\ref{B7}) in Eq.~(\ref{B1}) as compared to other terms.

Using Eqs.~(\ref{B5}) and (\ref{B6}), one obtains from Eq.~{B1})
\begin{equation}
\left.\frac{\partial G}{\partial\zeta_l^2}
\right\vert_{\zeta_l=0}=\frac{k_BT}{2\ho}\deE\left(
\frac{b_2}{y}+\frac{2b_1}{y^3}\right).
\label{B8}
\end{equation}

Substituting this equation to the second expression in Eq.~(\ref{eq75}),
we find
\begin{equation}
J_2=\frac{k_BT}{2\ho}\deE\sum_{l=1}^{\infty}
\int_{\zeta_l}^{\infty}\!\!\!dye^{-y}
\left(
\frac{b_2}{y}+\frac{2b_1}{y^3}\right)\zeta_l^2+\ldots\,.
\label{B9}
\end{equation}

Introducing the integration variable $v=y/\zeta_l$, one obtains from Eq.~(\ref{B9})
\begin{widetext}
\begin{eqnarray}
&&
J_2=\frac{k_BT}{2\ho}\deE
\int_{1}^{\infty}\!\!\!dv\left[\frac{b_2}{v}\sum_{l=1}^{\infty}\zeta_l^2e^{-v\zeta_l}
+\frac{2b_1}{v^3}\sum_{l=1}^{\infty}e^{-v\zeta_l}\right]+\ldots
\label{B10}\\
&&~~
=\frac{2k_BT}{\ho}\deE\left[b_2\tau^2\int_{1}^{\infty}\!
\frac{dv e^{\tau v}(e^{\tau v}+1)}{v(e^{\tau v}-1)^3}+
2b_1 \int_{1}^{\infty}\!
\frac{dv}{v^3(e^{\tau v}-1)}\right]+\ldots
\nonumber \\
&&~~
=\frac{k_BT}{\ho}\deE\frac{b_1+b_2}{\tau}\int_{1}^{\infty}\!
\frac{dv}{v^4}+\ldots
=\frac{k_BT}{\ho}\deE\frac{b_1+b_2}{3\tau}+\ldots\sim\deE
+\ldots\,.
\nonumber
\end{eqnarray}
\end{widetext}

Thus, a summation in nonzero $l$ again results in the additional factor
$\sim 1/\tau$. The same holds for all expansion terms in
the higher powers of $\zeta_l$ notated by dots in Eq.~(\ref{B10}).
Thus, $J_2$ contains the same exponentially decreasing with $T$
factor as $J_1$, and the result (\ref{eq78}) remains valid with account
of $J_2$.

%%%%%%%%%%%%%%%%%%%%%%%%%%%%%%%%%%%%%%%%%%%%%%%%%

%%%%%%%%%%%%%%%%%%%%%%%%%%%%%%%%%%%%%%%%%%%%%%%%%
%\end{document}
\newpage
%%%%%%%%
%%%%%%%%%%%%%
\begin{table*}
\caption{\label{tab1} Up to an order of magnitude asymptotic behaviors at
low temperature for the magnitudes of three different terms in the thermal
correction to the Casimir-Polder energy (columns 2--4), thermal correction
itself (column 5), and the Casimir-Polder entropy (column 6) for different
relationships between the energy gap and the chemical potential (column 1).
See the text for further discussion. }
\begin{ruledtabular}
\begin{tabular}{cccccc}
$\Delta$  &$r_{\alpha}(\ri\zeta_l,y,0)$&
 \multicolumn{2}{c}{$\dT r_{\alpha}(\ri\zeta_l,y,T)$}    &$r_{\alpha}(\ri\zeta_l,y,T)$ &\\
 versus  &$l\geqslant 0$& $l=0$& $l\geqslant 1$ & $l\geqslant 0$ & $S(a,T)$ \\
$\mu$ & $|\daT\cF|$ &$|\dbT\cF|$ & $|\dcT\cF|$ &$|\dT\cF|$ &\\ \hline
&&&&&\\[-6mm]
$\Delta>2\mu$ & $\frac{\alpha_0(k_BT)^5}{(\hbar c)^3\Delta}$ &
$\frac{\alpha_0(k_BT)^2}{\hbar c a^2}e^{-\frac{\Delta-2\mu}{2k_BT}}$ &
$\frac{\alpha_0 k_BT}{ a^3}e^{-\frac{\Delta-2\mu}{2k_BT}}$ &
$\frac{\alpha_0(k_BT)^5}{(\hbar c)^3\Delta}$ &
$\frac{\alpha_0 k_B(k_BT)^4}{(\hbar c)^3\Delta}$ \\[2mm]
$\Delta<2\mu$ & $\frac{\alpha_0\mu^2(k_BT)^2}{(\hbar c)^2 a
(4\mu^2-\Delta^2)^{1/2}}$ &
$\frac{\alpha_0 k_BT}{a^3}e^{-\frac{2\mu-\Delta}{2k_BT}}$ &
$\frac{\alpha_0 \hbar c}{ a^4}e^{-\frac{2\mu-\Delta}{2k_BT}}$ &
$\frac{\alpha_0\mu^2(k_BT)^2}{(\hbar c)^2 a
(4\mu^2-\Delta^2)^{1/2}}$ &
$\frac{\alpha_0\mu^2k_B^2 T}{(\hbar c)^2 a
(4\mu^2-\Delta^2)^{1/2}}$  \\[2mm]
$\Delta=2\mu$ & $\frac{\alpha_0(k_BT)^5}{(\hbar c)^3\Delta}$ &
$\frac{\alpha_0(k_BT)^2}{\hbar c a^2}$ &
$\frac{\alpha_0 k_BT}{ a^3}$ &
$\frac{\alpha_0 k_BT}{ a^3}$ &
$\frac{\alpha_0 k_B}{ a^3}$
\end{tabular}
\end{ruledtabular}
\end{table*}
\end{document}